\documentclass[a4paper]{jpconf}
\usepackage{graphicx}
\def\aj{AJ}%
\def\araa{ARA\&A}%
\def\apj{ApJ}%
\def\aap{A\&A}%
\def\aaps{A\&AS}%
\def\mnras{MNRAS}%
\def\pasp{PASP}%
\def\mathstacksym#1#2#3#4#5{\def#1{\mathrel{\hbox to 0pt{\lower#5\hbox{#3}\hss} \raise #4\hbox{#2}}}}

\mathstacksym\gta{$>$}{$\sim$}{1.5pt}{3.5pt} % greater than approximately
\mathstacksym\lta{$<$}{$\sim$}{1.5pt}{3.5pt} % less than approximatel

\begin{document}
\title{Towards $\mu$-arcsecond spatial resolution with Air Cherenkov Telescope arrays as
  optical intensity interferometers}

\author{W J  de Wit$^1$, S  LeBohec$^2$, J A  Hinton$^1$, R J  White$^1$
  M K  Daniel$^3$ and J Holder$^4$}
\address{$^1$ School of Physics \& Astronomy, University of Leeds, LS2 9JT,
  Leeds, UK}
\address{$^2$ Department of Physics, University of Utah, 115 S 1400 E, Salt Lake  City, UT 84112-0830, USA}
\address{$^3$ Department of Physics, Durham University, South Road, Durham, DH1 3LE, UK}
\address{$^4$ Department of Physics and Astronomy, University of Delaware,
  Newark Delaware, USA}

\ead{w.j.m.dewit@leeds.ac.uk}

\begin{abstract}
In this poster contribution we highlight the equivalence between an Imaging
Air Cherenkov Telescope (IACT) array and an Intensity Interferometer for a range
of technical requirements. We touch on the differences between a Michelson
and an Intensity Interferometer and give a brief overview of 
the current IACT arrays, their upgrades and next generation concepts 
(CTA, AGIS, completion 2015). The latter are foreseen to include 30-90 telescopes that will provide 400-4000 different
baselines that range in length between 50\,m and a kilometre. Intensity
interferometry with such arrays of telescopes attains 50$\mu$-arcseconds
resolution for a limiting $\rm m_{v}\sim8.5$. This technique opens the
possibility of a wide range of studies, amongst others, probing the stellar
surface activity and the dynamic AU scale circumstellar environment of stars
in various crucial evolutionary stages. Here we discuss possibilities for
using IACT arrays as optical Intensity Interferometers.

\end{abstract}

\section{Introduction}
Over the past decade, Michelson stellar interferometry has seen some
tremendous advances in applicability. It has evolved from a primarily
experimental technique towards a general astrophysical observational mode
extensively used by the community for galactic and extragalactic science (see
the various contributions in these proceedings). Observations with milli
arcseconds angular resolution are now routinely performed in the near
infrared. Strategic planning of next generation Michelson interferometers
should aim for higher resolution, higher sensitivity and image reconstruction
capabilities (synthesis or direct imaging). An alternative technique to
Michelson interferometry that has the potential of delivering similar
scientific products is the technique of intensity interferometry (II).  It is
able to attain $\mu$-arcseconds resolution and provide image synthesis capacity
and can be implemented in a cost effective and straightforward instrument.

The principle of II is based on the partial correlation of
intensity fluctuations of coherent light beams measured at different points in
space or in time. The fluctuations (also called wave-noise, although
``noise'' is a bit of a misnomer) was a well-known phenomenon at radio
wavelengths, but its nature and even reality at optical wavelengths were
seriously debated in the 1950s. Hanbury Brown \& Twiss \cite{1956Natur.178.1046H,1958RSPSA.248..199B} conclusively
demonstrated in a series of papers how the phenomenon is firmly rooted in both
theory and experiment. The intensity fluctuations can be interpreted in a semi-classical sense as the
superposition of light waves at different frequencies producing beats. The
fluctuations can also be interpreted in a quantum mechanical sense as an
effect related to so-called photon bunching \cite{clarkjones53}. At optical wavelengths, the resulting fluctuations are much smaller than
the shot-noise component of a (recorded) light beam. Given that shot-noise is
random, any coherence of different light beams is
solely determined by the intensity fluctuation effect. 
Hanbury Brown et al. \cite{1974MNRAS.167..121H} measured the stellar
diameter for 32 stars using a dedicated stellar intensity interferometer
stationed at Narrabri, Australia. Their pioneering experiments constituted the first
successful stellar diameter measurements after the Michelson \& Pease
experiments. 
%It is worthwhile to stress that intensity interferometry and
%Michelson interferometry rely on two different physical phenomena, each with
%its pratical advantages and disadvantages, which we will highlighten in the
%following sections. All modern interferometres, like the VLTI and Keck-I are
%based on the principle of Michelson stellar interferometry.  
We will discuss in this contribution the potential of future Cherenkov Telescope Arrays for
a revival of intensity interferometry as a mainstream high angular resolution
imaging technique in astronomy.

\section{Basic differences between Michelson and Intensity interferometry}
All modern interferometers in operation are based on the principle of
Michelson stellar interferometry. It provides a measure of the spatial
coherence of two light beams by using the fringe contrast (visibility)
of the beams' interference pattern.  The technique requires highly
accurate metrology to correct for path differences between the beams,
which is on the order of the wavelength of light itself. Active
correction of the optics (adaptive optics, tip-tilt) are required in
order to make the otherwise corrugated wave front planar. Light beams
are made to interfere, and allow in principle the determination of the
modulus and phase of the complex visibility. In practise however,
atmospheric limitations corrupt the phase (the fringes wander over the
detector), and in broad-band interferometric observation a quadratic
visibility estimator ($VV^{*}$) is generally employed eliminating the
corrupted phase, but at the same time introducing a bias in the
estimator that needs to be taken care of,
e.g., \cite{1999PASP..111..111C,1991AJ....101.2207M}. In spectrally
resolved observations a linear visibility estimator can be employed,
thanks to the achromatic nature of the atmospheric disturbances, and
relative phase can be retrieved. Absolute phase information with
ground based interferometers can be obtained by simultaneously
observing a phase reference star \cite{2008NewAR..52..199D}, or by
exploiting the principle of closure phase when working with more than
two telescopes, see \cite{2000plbs.conf..203M}.

Intensity interferometry provides a measure of the square of the Michelson 
fringe visibility. This fundamental difference is the basis for some of
its technical advantages. II is virtually insensitive to atmospheric and
instrumental instabilities as no actual waves are interfered. The technique 
can thus make do with relatively coarse light collectors and long 
baselines. The signals from each receiver can be electronically recorded and
correlated after detection. The correlation of the multiple signals
measured by a telescope array can thus all be determined for all possible
pairs, and maybe even higher order correlations can be exploited. The
coherence length is set by the length of the frequency beats (wave
group) which is of order cm rather than nm (depending on bandwith), alleviating
the strong constraints on accurate delay tracking.

II is however less sensitive in measuring coherence than Michelson
interferometry, as the sought fluctuations are very small relative to
the shot-noise of the photon stream.  However, without stringent requirements
for the light collectors, the decreased sensitivity can be offset
by large coarse mirrors in order to maximise photon
collection. In addition \cite{2006MNRAS.368.1652O,2006MNRAS.368.1646O}
have shown that under certain conditions higher order correlations
will increase the sensitivity (see \cite{lebohec2008}). A second
drawback is no phase information is contained within the
measurement. Although this does not impede the measurement of
centro-symmetric objects (indeed as was done with the Narrabri Stellar
Intensity Interferometer), it has been shown that the phase can be
recovered using correlations between more than two signals
\cite{1963JAP....34..875G,2004JOSAA..21..697H}. Also superimposing a coherent beam from a
known reference source on the light beam of the target source would
allow the recovery of the phase \cite{1963JAP....34..875G}, see also
\cite{1968JOSA...58.1233M,1969ApPhL..15..227B}.

%``The signals are correlated after detection. Delay tracking and local oscillator stability don't have to be as
%accurate'', from weisstein World o fphysics.
%First order and second order correlations (square degree)

\section{IACT array as an optical intensity interferometer}
Imaging Air Cherenkov Telescope (IACT) Arrays are multi-telescope arrays designed to image air
showers that are produced by high energy particles and $\gamma$-ray photons
($>10$ GeV) that impinge on the earth's atmosphere. The air showers consist of
secondary particles some of which carry electrical charge and produce 
Cherenkov light. As the Cherenkov radiation is faint (depending on the energy
of the incident particle/photon), large collecting areas are required to obtain
decent signal-to-noise ratios. The flash of Cherenkov radiation produced is also very
brief (a few nanoseconds) and fast photon counting detectors are therefore
mandatory. The advantage of employing telescope arrays in air-shower
observations is the ability to reconstruct the spatial geometry of the shower
in the earth's atmosphere. This allows one to make a distinction between
cosmic ray showers and $\gamma$-ray showers, and also to find the spatial
direction of the incident particle/photon, thus obtaining angular information
on the galactic or extragalactic source. In short, IACTs aim to
determine the energy and the astrophysical source of $\gamma$-ray photons by observing the 
optical light of induced Cherenkov radiation. It should be clear that the basic technical specifications
of an IACT array as touched upon here are very similar to the ones of an Intensity
Interferometer as briefly described in the previous section. A more detailed
description of the components of IACT arrays relevant for II (fast photon detection, signal
communication, correlators) can be found in \cite{2006ApJ...649..399L}.

Currently, there are two major and successful IACT arrays
operational. The H.E.S.S. array \cite{2004NewAR..48..331H} consists of
4 telescopes of 108\,m$^2$ each. They are arranged in a square with
120\,m side length. The VERITAS \cite{2006APh....25..391H} array is
similar and also consists of 4 telescope with 110\,m$^2$ light
collecting area. In contrast to HESS, it has variable baselines
ranging from 34 to 109 metres. HESS will be upgraded to HESS II with
one extra 30\,m dish telescope placed at the centre of the
configuration, its completion foreseen for the 4th quarter of 2009. Future IACT projects are currently under study, such as
AGIS\footnote{\tt{http://gamma1.astro.ucla.edu/agis/}} and
CTA\footnote{\tt{http://www.mpi-hd.mpg.de/hfm/CTA}}, and aim at
increasing the number of telescopes up to 100, each with a
$\sim100$\,m$^2$ light collectors arranged over an area covering
1km$^2$ or more. These envisioned projects (expected to be operational around 2015)
would not only benefit the resolving power and
sensitivity for the high-energy science purposes of IACTs, but they
are in harmony with the requirements of a more sensitive intensity
interferometer with a higher angular resolution and denser
(u,v)-coverage \cite{2006ApJ...649..399L}. Future IACTs will offer
thousands of baselines from 50\,m to at least 1\,km and when used as
an intensity interferometer may attain angular resolution down to a
few tens of $\mu$-arcseconds.

\begin{figure}
  \begin{center}
    \includegraphics[width=6cm,height=6cm]{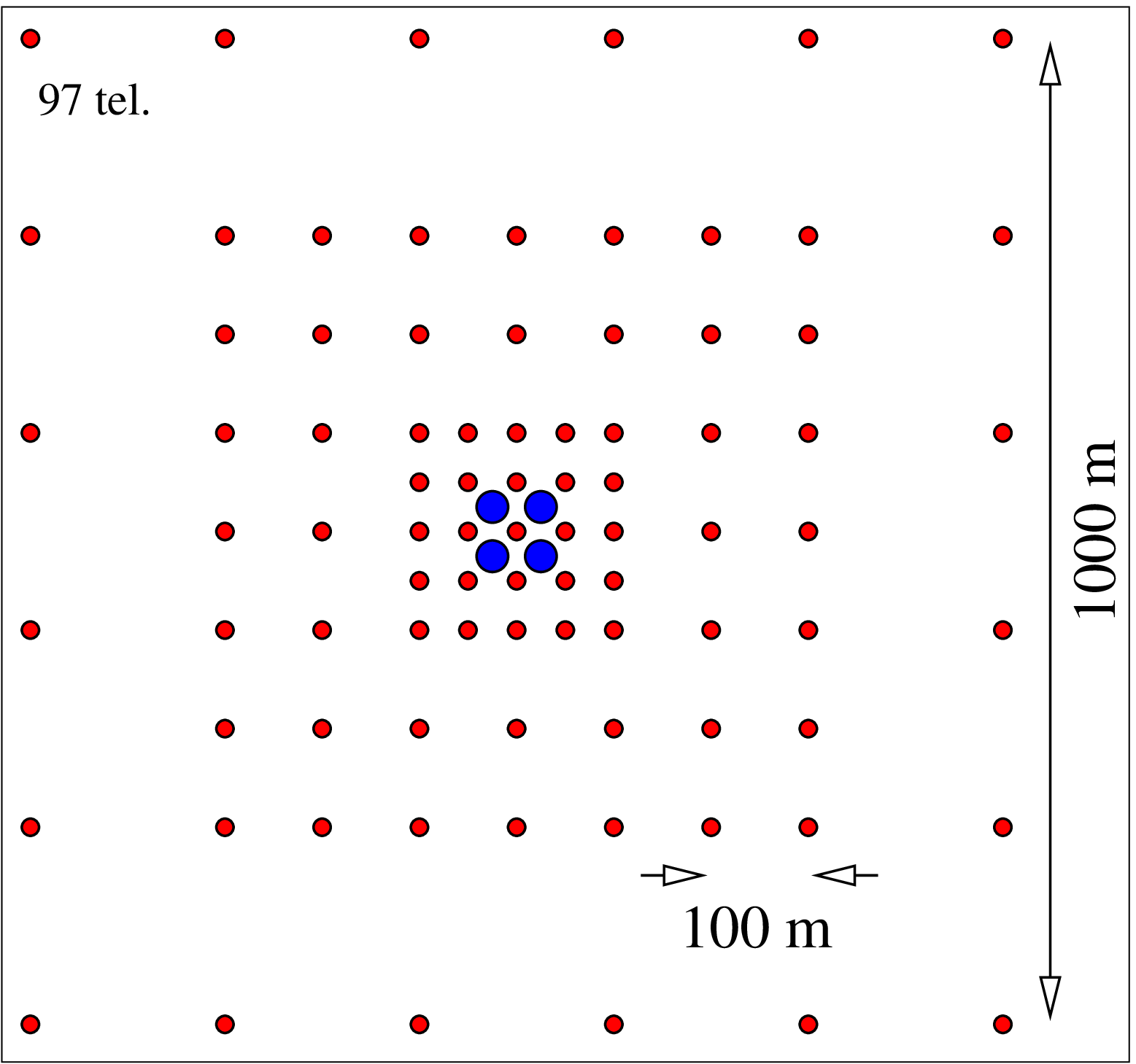}
    \includegraphics[width=8cm,height=6cm]{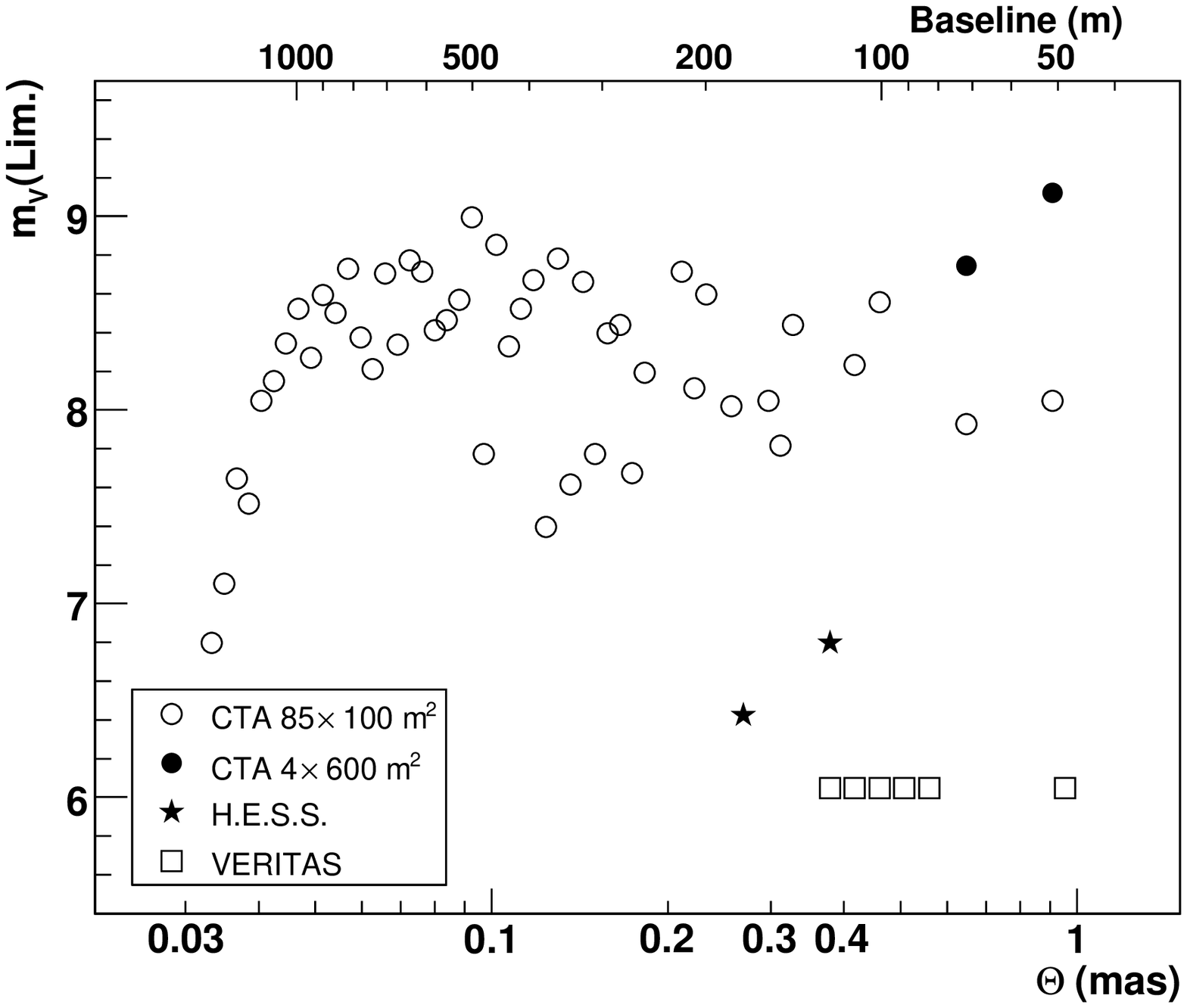}
  \end{center}
  \caption{\label{fig1}{\it Left:} Proposed lay-out for the future CTA. Small
red dots are the 85 $\rm 100\,m^2$ dishes, large blue dots are the four $\rm
100\,m^2$ dishes (adapted from \cite{bernloehr}). {\it Right:} Sensitivity estimate as function of
the 48 non-redundant baselines for a 5$\sigma$ detection, in a 5 hr
integration on a centro-symmetric object with 50\% visibility. Final values depend on signal bandwith
and CTA design details, see \cite{2006ApJ...649..399L}.}
\end{figure}

\section{Science for II}
Limiting $\rm m_{v}$ of the CTA concept is illustrated in the right panel of
Fig.\,1 with simulated performance that meet the goal sensitivity for
$\gamma$-ray astronomy \cite{bernloehr}. Targets are limited to a $\rm
m_{v}\approx8.5^{m}$ for a S/N = 5, and a 5 hours integration in case of
50\% visibility (see \cite{lebohec2008}). These specifications
allow important interferometry studies regarding binary stars, stellar radii and
pulsating stars with unprecedented resolution on $\mu$-arcsecond scales. We
highlight three potential science cases below.

\paragraph{\bf{Star formation}}
Key questions relating to the physics of mass accretion and pre-main sequence
(PMS) evolution can be addressed by means of very high resolution imaging as
provided by the next generation IACTs and II. They involve the
absolute calibration of PMS tracks, the mass accretion process, continuum
emission variability, and stellar magnetic activity. The technique may allow 
us to the resolve spot features on the stellar surface.  Hot spots deliver
direct information regarding the accretion of material onto the stellar
surface. Cool spots, on the other hand, may cover 50\% of the stellar surface,
and they are the product of the slowly decaying rapid rotation of young
stars. Imaging them will constrain ideas regarding the interplay of rotation,
convection, and chromospheric activity as traced by cool spots. It may provide
direct practical application as the explanation for the anomalous photometry
observed in young stars \cite{2003AJ....126..833S}.

In practise, about 50 young stars with $\rm m_{v}<8^{\rm m}$ are within reach of future
IACTs. In the last decade several young coeval stellar groups have been
discovered in close proximity ($\sim$50\,pc) to the sun. Famous
examples are the TW Hydra and $\beta$ Pic comoving groups. The majority of the spectral types
within reach range between A and G-type. Their ages lie within the range 8
to 50 Myr (see \cite{2004ARA&A..42..685Z} for an overview).  The age
intervals ensures that a substantial fraction of the low-mass members are still
in their PMS contraction phase. Measurement of their angular size can be used in
the calibration of evolutionary tracks, fundamental in deriving the properties
of star forming regions and young stellar clusters.  The proximity
of the comoving groups ensures that their members are bright. Their proximity renders the
comoving group also relatively sparse making them very suitable, unconfused targets 
despite the large optical PSF of a few arcminutes. The sparseness is also the
reason for incomplete group memberships, making it likely that the number of
young stars close to the sun will increase with the years to come.

%Disk truncation radius close to the co-rotation radius of 0.1
%AU. Accretion-ejection.

%Veiling: In Herbst \& Shevchenko (1999) there are only few variable TTauri
%stars with magnitude less than V=9m. There are about 10 HAeBe stars.

%Transition planets
\paragraph{\bf{Distance scale and pulsating stars}}
Measuring diameters of Cepheids is a basic method with far reaching
implications. A radius estimate of a Cepheid can be obtained using the
Baade-Wesselink method. The Baade-Wesselink method relies on the measurement of
the ratio of the star size at times $t_1$ and $t_2$, based on the luminosity and
colour. Combined with a simultaneous measurement of the radial velocity, this method
delivers the difference in the radius between $t_1$ and $t_2$. With the known
difference and ratio of the radius at two times, one can derive the radius of
the Cepheid. Combining II angular size measurement with the radius estimate
one obtains the distance to the Cepheid (see \cite{1994ApJ...432..367S}). This
makes possible the calibration of the all important Cepheid period-luminosity
relation using local Cepheids. A count of Cepheids observed with {\it Hipparcos}
\cite{1999A&AS..139..245G} shows that at least 60 Cepheids with $\rm m_{v}
<8^{m}$ are in reach with future IACTs.

\paragraph{\bf {Rapidly rotating stars}}
As a group, classical Be stars are particularly well-known for their close to
break-up rotational velocities as deduced from photospheric absorption lines.
In addition the stars show Balmer line emission firmly established as due to
gaseous circumstellar disks, that appear and disappear on timescales of months
to years. These two properties are somehow related, but many open questions
regarding the detailed physical processes at play exist.

The Be-phenomenon is an important phenomenon given the number of stars and
stellar physics involved (fraction of Be stars to normal B-type peaks
at nearly 50\% for B0 stars, \cite{1997A&A...318..443Z}). Absorption lines will
however never provide the final answer to their actual rotational velocity due
to strong gravity darkening at the equator and brightening at the pole
areas. Direct measurement of the shape of the rotating star is not hampered by
gravity darkening, and provides a direct indication of the rotational speed (see,
e.g., $\alpha$ Eri with the VLTI, \cite{2003A&A...407L..47D}).
The Be star disk formation and dissolution activity is little
understood. Photometric observations of Be star disks seem to indicate that they
may actually evolve into ring structures before disappearing into the
interstellar medium (e.g. \cite{2006A&A...456.1027D}). The disk's Bremsstrahlung
can constitute $\sim 30\%$ of the total light in $V$-band
\cite{1997A&A...318..443Z}.

There are about 300 Be stars\footnote{\tt{http://www.astrosurf.com/buil/us/becat.htm}} brighter than $m_{v}=8^m$, roughly
corresponding to a distance limit of 700\,pc. Signifying that Be star phenomena
can be probed in depth with IACT based II.

\section{Concluding remarks}
Technical requirements for a relatively sensitive and $\mu$-arcsecond resolution
synthesis intensity interferometer go hand in hand with the designs for next
generation Air Cherenkov Telescope arrays. This prompts close study of the
possibilities in designing and building a IACT instrument that incorporates
the II capability. In principle there is no competition between the two modes, as
$\gamma$-ray observations need to be executed when the moon is less than half
full. This leaves half the available night time of an IACT for interferometric 
observations.

The renewed interest in II has resulted in the formation of an IAU working
group on intensity interferometry. Laboratory experiments are performed to
test and demonstrate the various aspects of II integrated in a IACT. A pair of
3-m telescopes is now available in Utah, USA (Star base Utah 2008\footnote{\tt{http://www.physics.utah.edu/\,$\tilde{ }$\,lebohec/StarBaseWeb/}}) for testing the techniques in a realistic astronomical environment.  

In conclusion, the time is now to assess the applicability of II as a future
high resolution imaging mode that allows science to enter the
realm of astrophysics on $\mu$-arcsecond angular scales.

\ack
This contribution is partly based on two presentations \cite{2008AIPC..984..205L,2008AIPC..984..268D} given at the meeting 
``The Universe at Sub-Second Timescales'' held in Edinburgh, September 2007.

\section*{References}
%\bibliography{II.bib}
\providecommand{\newblock}{}

\end{document}